\documentstyle[12pt,epsf]{article}
\newcommand{\beqn}{\begin{eqnarray}}
\newcommand{\eeqn}{\end{eqnarray}}
\newcommand{\beqns}{\begin{eqnarray*}}
\newcommand{\eeqns}{\end{eqnarray*}}

\newcommand{\rmi}{\mbox{i}}

\newcommand{\pdd}[2]{{\partial{#1}\over\partial{#2}}}

\newcommand{\llaa}{\langle\!\langle}
\newcommand{\rraa}{\rangle\!\rangle}

\newcommand{\tr}{\mathop{\mbox{Tr}}}

\newcommand{\be}{\begin{eqnarray}}
\newcommand{\ee}{\end{eqnarray}}

\begin{document}

\gdef\journal#1, #2, #3, 1#4#5#6{{#1} {\bf #2} (1#4#5#6) #3}
\gdef\ibid#1, #2, 1#3#4#5{{#1} (1#3#4#5) #2}

\begin{flushright} Oslo SHS-96-9, OSLO-TP 4-97 \\
hep-th/9702066
\end{flushright}

{\centerline{\Large Algebra of Observables for Identical Particles in
One
Dimension}}
\vskip 1cm

{\centerline {Serguei B. Isakov$^{a,b}$, Jon Magne Leinaas$^{a,b}$,
Jan Myrheim$^{a,c}$,}}
{\centerline {Alexios P. Polychronakos$^{a,d}$, Raimund
Varnhagen$^{a,e}$}}

\medskip
{\centerline{\it ${}^{a}$Senter for H\o yere Studier, 
Drammensveien 78, 0271 Oslo, Norway }}
{\centerline{\it ${}^{b}$Department of Physics,  University of Oslo,
P.O. Box 1048 Blindern, N-0316 Oslo, Norway}}
{\centerline{\it ${}^{c}$Department of Physics, University of 
Trondheim,
N--7034 Trondheim, Norway}}
{\centerline{\it ${}^{d}$Theoretical Physics Department, Uppsala
University,
Box 803, S--75108 Uppsala, Sweden }}
{\centerline{\it ${}^{e}$Blaupunktwerke GmbH,
K7/EFB1, 
Postfach 77 77 77, 
D-31312 Hildesheim,
Germany}}

\date{\today}

%\maketitle

 \begin{abstract}
The algebra of observables for
identical particles on a line
is formulated starting from postulated basic commutation relations.
A realization of this algebra in the Calogero model
was previously known. New realizations are presented here
in terms of differentiation operators and in terms of
$\mbox{SU}(N)$-invariant
observables of the Hermitian matrix models.
Some particular structure properties of the algebra are briefly
discussed.

\end{abstract}

\section{Introduction}

The idea of the Heisenberg quantization of identical particles is to
identify a fundamental {\em algebra of observables} where
the elements do not refer to individual particles, but are symmetric 
with
respect to particle indices
\cite{WHPAMD,LM-Heis}.
Quantization amounts to choosing an 
irreducible representation of the algebra. The
restriction to symmetric variables, in general allows for the
existence of representations which are not present when non-symmetric 
variables are also
included. This gives an algebraic way to introduce generalized
particle statistics, where different statistics correspond to
inequivalent representations of the same fundamental algebra.

For two identical (spinless) particles on a line, all the
observables for the relative motion can be generated
from the quadratic observables in the relative position $x$
and momentum $p$,
which form the  algebra $sl(2,{\bf R})$ \cite{LM-Heis}.
The irreducible representations of the same
algebra also classify the solutions of the Calogero model \cite{Cal} 
for two particles
\cite{Perelomov}, and the Heisenberg quantization thus suggests the
interpretation of
the singular $1/x^2$-potential of the Calogero model as a
``statistics interaction''
between the particles. This gives a specific way to introduce
{\it fractional statistics in one dimension} \cite{LM-Heis} (see also
discussion of the many-body case in terms of
the Schr{\"o}dinger quantization in Ref.~\cite{PolyNPB}).
The same algebra --- and, correspondingly, the notion of fractional
statistics in one dimension --- applies to anyons restricted to the
lowest
Landau level, where the dynamics of particles becomes effectively
one-dimensional
\cite{HLM}.

Note that these algebraic arguments for the $1/x^2$ potential as a
statistics
interaction have found support in
statistical  mechanics. The statistical distribution for fractional
statistics in one  dimension defined in the above way
\cite{I-IJMPA} is the same as that for a more recent alternative
definition  of fractional statistics, the so  called {\it exclusion}
statistics \cite{Haldane}, which is introduced in statistical 
mechanical
terms (see \cite{I-MPLB,Wu94}).
The thermodynamic quantities of anyons in the lowest Landau
level
\cite{dVO94} represent the same
statistical mechanics.

For the Calogero model it was shown some time ago, by use of the 
so-called exchange
operator formalism, that the quadratic variables in $x$ and $p$ can be
expressed in
terms of linear (non-observable) variables in a way that the
representations of
$sl(2,{\bf R})$ corresponding to fractional statistics be reproduced. 
This amounted to
introducing a modification of the fundamental commutator between (the 
relative)
position and momentum. It was also shown that this generalized 
commutation
relation could be extended to a set of commutation relations for the 
general
$N$-particle system. This modified algebraic structure has been 
referred to
as an
$S_N$-{\em extended} Heisenberg algebra since it includes also 
permutation
operators for the particles. A typical feature of the modified 
Heisenberg
commutation
relations is that they involve non-symmetric operators in
$x$ and
$p$ and depend explicitly on the (statistics) interaction
parameter of the Calogero model
\cite{Poly-PRL92,BHV}.

Motivated by these results, in Ref.~\cite{G} the question was studied
whether it is possible, starting from the parameter-dependent
$S_N$-extended Heisenberg algebra, to construct a closed algebra
of symmetric one-particle operators which would play the  role of the
algebra  of observables for more than two particles. It was shown that
such an algebra (referred to as ${\cal G}$) can indeed be 
constructed. It is
an infinite dimensional Lie algebra which is independent of the 
statistics
interaction parameter \cite{G}. (One should note ${\cal G}$ is a larger
algebra than the related algebra
$W_{1+\infty}$.) In this Letter we show how the algebra ${\cal G}$
can naturally be incorporated into the Heisenberg quantization
scheme, starting in Sect.~2 with a discussion of the one-particle
observables for identical particles in one dimension.

In Ref.~\cite{G}
the explicit construction of the elements of the algebra
${\cal G}$  was based on the  $S_N$-extended Heisenberg algebra, 
where the
defining relations involve both single-particle and  two-particle 
operators
and depend on the statistics parameter. However, since the
commutation relations of the algebra ${\cal G}$ (which in fact 
{\em define}
the algebra)  involve  only one-particle operators and do not contain 
the
statistics  parameter, it was natural to expect that there should 
exist a
simpler way to  formulate this algebra.  In Sect.~3 we give such a
formulation starting from a set of {\em basic} commutation relations 
which
are more general than the canonical commutation relations and which 
do not
depend on the statistics parameter. This formulation allows one to 
find new
realizations of the algebra ${\cal G}$. In Sect.~4 we give a brief
discussion of three different representations, that of the Calogero
model, a representation in the form of differentiation operators and
finally a representation referring to the Hermitian matrix models.  The
latter two realizations
are in fact much simpler for
studying the structure of the algebra than the original one \cite{G}. 
 In
Sect.~5 we present some results of computer calculations concerning the
number of independent elements of the algebra for low-order 
polynomials in
$x$ and $p$.

%%%%%%%%%%%%%%%%%%%%%%%%%%%%%%%%%%
\section{Heisenberg quantization}
%%%%%%%%%%%%%%%%%%%%%%%%%%%%%%%%%%

A system of $N$ identical particles on the line is described
classically by $N$ points $(x_1,p_1)$, $(x_2,p_2),\ldots,(x_N,p_N)$
in the one-particle phase space, taken in arbitrary order.
Every observable must be a symmetric function of the $N$ points
\cite{WHPAMD}.
By definition, a set of symmetric functions is {\em complete} if it
contains a set of (at least) $2N$ functions such that the $2N$ (or 
more)
function values determine the $N$ one-particle phase space points
uniquely up to permutations. A complete set
of more than $2N$ functions is overcomplete.

One particular complete set of symmetric functions is the set of all
one-particle observables.  By definition, a
one-particle observable is
given by an observable $f=f(x,p)$ of the one-particle system as
\beqn
\llaa f\rraa=N\langle f\rangle=\sum_{j=1}^N f(x_j,p_j)\;,
\eeqn
where $\langle f\rangle$ is the average of $f$ over the $N$ particles.
We may write more compactly $\llaa f\rraa=\sum_j f_j$.
We call $f$ a {\em single-particle
function} (``non-observable''),
as opposed to
the {\em one-particle observable} $\llaa f\rraa$.

An important property of the set of one-particle observables
is that it is closed under Poisson brackets, in the classical case.
The mapping $f\mapsto\llaa f\rraa$ preserves Poisson brackets,
\beqn
\{\llaa f\rraa,\llaa g\rraa\}=\sum_j \{f_j,g_j\}=\llaa\{f,g\}\rraa\;.
\eeqn
Similar relations hold in the standard quantization, based on
the canonical Heisenberg commutation relations
\beqn
\label{eq:Heisenbergalg}
{[x_j,p_k]}=\rmi\hbar\delta_{jk}\;,\qquad
{[x_j,x_k]=[p_j,p_k]}=0\;.
\eeqn

For any given $N$ there exists a minimal complete set consisting of
exactly $2N$
one-particle classical observables. For example,
choose the symmetric and homogeneous polynomials
\beqn
\llaa x^m\rraa=\sum_j {x_j}^m\;,\qquad
\llaa x^np\rraa=\sum_j {x_j}^np_j\;,
\eeqn
with $m,n=1,2,\ldots,N$. This set is closed under Poisson brackets.
However, it makes a fundamental distinction between
coordinates and momenta,
since the momentum dependence is at most linear.

The algebra to be considered here is
an extension which is overcomplete, but
treats coordinates and momenta on an equal footing. The
commutation relations of the algebra are generic in the sense of 
being valid
for arbitrary particle number $N$. A number of algebraic identities  
are
necessary consequences of the commutation relations. When $N$ is fixed
there will be additional algebraic relations between one-particle
observables, but these $N$-dependent relations are not considered to be
characteristic for the algebra itself.

%%%%%%%%%%%%%%%%%%%%%%%%%%%%%%%%%%%%%%
\section{Definition of the algebra
${\cal G}$
}
\label{definition}
%%%%%%%%%%%%%%%%%%%%%%%%%%%%%%%%%%%%%%

The Heisenberg commutation relations,
or the corresponding classical Poisson bracket relations,
are too detailed, since they
involve operators,
or phase space functions,
that are not symmetric and therefore not observable.
We want to consider here commutation relations that can be deduced
from the Heisenberg relations, but are less detailed and therefore
allow more general quantum theories.
%In particular, these weaker
%relations can not be used in the same way as the Heisenberg
%relations for reordering operator products of $x_j$ and $p_j$.
%It will not even be true in general that the commutator of two
%one-particle operators is a one-particle operator, although
%it is true for the particular algebra that we will discuss.

The algebra we want is generated from the set of observables
$\llaa x^m\rraa$ and $\llaa p^n\rraa$, with $m,n=1,2,\ldots$,
by means of Poisson brackets in the classical case and commutators
in the quantum case. We postulate the following {\em basic} commutation
relations, which follow directly from the Heisenberg relations,
eq.~(\ref{eq:Heisenbergalg}),
but are genuinely weaker,
\beqn
\label{eq:genHeisenbergalg}
{[\llaa x^m\rraa,p_j]}=\rmi m\hbar{x_j}^{m-1}\;,\quad
{[\llaa p^n\rraa,x_j]}=-\rmi n\hbar{p_j}^{n-1}\;,\quad
{[\llaa x^m\rraa,x_j]}={[\llaa p^n\rraa,p_j]}=0\;.
\eeqn
The
three examples given in Sect.~4  show that
there exist non-trivial realizations.

Note that if $A$ is one of the operators $\llaa x^m\rraa$,
$\llaa p^n\rraa$, and if $\llaa f\rraa$ is any one-particle
operator, then the commutator
\beqn
{[A,\llaa f\rraa]}=\sum_j{[A,f_j]}
\eeqn
is a one-particle operator, by
eq.~(\ref{eq:genHeisenbergalg}).
For example,
\beqn
{[\llaa x^5\rraa,{x_j}^2{p_j}^3]}=5\rmi\hbar(
{x_j}^6{p_j}^2+{x_j}^2p_j{x_j}^4p_j+{x_j}^2{p_j}^2{x_j}^4)\;. \nonumber
\eeqn
Hence every operator of the form
\beqn
B_k={[A_k,\,[A_{k-1},\,[\,\ldots\,[A_2,A_1]\,\ldots\,]\,]\,]}
={[A_k,B_{k-1}]}\;,
\eeqn
where each $A_j$ is one of the $\llaa x^m\rraa$ or $\llaa p^n\rraa$
operators, is a one--particle operator. It follows by
repeated applications of the Jacobi identity, in the form
\beqn
{[B_k,C]}={[A_k,\,[B_{k-1},C]\,]}-{[B_{k-1},\,[A_k,C]\,]}\;,
\eeqn
that the vector space spanned by all operators of the form $B_k$
is closed under commutation and hence is a Lie algebra.
It is this Lie algebra
we want to study here, and we will call it ${\cal G}$.

Note that when we use eq.~(\ref{eq:genHeisenbergalg}) to compute
an explicit expression for a non-zero operator of the form $B_k$ as
a symmetric one-particle operator, we will in general get two different
results, because we may expand the innermost commutator ${[A_2,A_1]}$
in two different ways. Either as
\beqn
\label{xp1}
{[\llaa x^m\rraa,\llaa p^n\rraa]}=
\sum_j{[\llaa x^m\rraa,{p_j}^n]}=
\rmi m\hbar\sum_{\ell=0}^{n-1}
\llaa p^{\ell}x^{m-1}p^{n-1-\ell}\rraa\;,
\eeqn
or as
\beqn
\label{xp2}
{[\llaa x^m\rraa,\llaa p^n\rraa]}=
\sum_j{[{x_j}^m,\llaa p^n\rraa]}=
\rmi n\hbar\sum_{\ell=0}^{m-1}
\llaa x^{\ell}p^{n-1}x^{m-1-\ell}\rraa\;.
\eeqn
It is a consistency condition that the two expressions for $B_k$
must be equal.
Other consistency conditions follow from the Jacobi identity,
for example that
\beqn
{[\llaa x^k\rraa,\,[\llaa x^m\rraa,\llaa p^n\rraa]\,]}=
{[\llaa x^m\rraa,\,[\llaa x^k\rraa,\llaa p^n\rraa]\,]}\;.
\eeqn
In this way we get a number of identities which can
sometimes be used to reorder a product of operators $x_j$ and $p_j$
when there is a sum over the particle index $j$.

One should note that the basic commutation relations
(\ref{eq:genHeisenbergalg}) can be interpreted as giving an abstract
definition of the
Lie
algebra $ {\cal G}$ where the explicit expression of the
elements as sums over single-particle variables is not needed.  In this
abstract formulation an element $\llaa f\rraa$ is {\em defined} by its
commutators with $x_j$ and $p_j$. All elements of the algebra can be
constructed by repeated use of the fundamental commutators for $\llaa
x^m\rraa$ and $\llaa p^n\rraa$, and
an identity
between observables then simply means that they have identical
commutators with both $x_j$ and $p_j$.
This abstract definition is in itself a realization of the algebra
in terms of differentiation operators, as discussed in some more
detail in Subsect.~\ref{subsect:Diffop}.

%%%%%%%%%%%%%%%%%%%%%%%%%
\section{Realizations of
${\cal G}$}
\subsection{The $S_N$--extended Heisenberg algebra}
%%%%%%%%%%%%%%%%%%%%%%%%%%%%%

Our first example is the extended $N$-particle Heisenberg
algebra defined by the following commutation relations,
containing the arbitrary
real
parameter $\lambda$,
\beqn
\label{eq:basrel1}
{[x_j,x_k]}={[p_j,p_k]}=0\;,\qquad
{[x_j,p_k]}=
\left\{\begin{array}{lll}
-\rmi\lambda\hbar K_{jk} & \mbox{if} & j\neq k\;,\\
\rmi\hbar+\rmi\lambda\hbar
\mathop{\sum}_{\ell\neq k}K_{\ell k} & \mbox{if} & j=k\;.
\end{array}\right.
\eeqn
The operators
$K_{jk}=K_{jk}^{\dag}=K_{kj}$
%))) CHANGE 970107 JM
are defined for $j\neq k$, and
satisfy the following relations
when no two of the indices $j,k,\ell,m$ are equal,
\beqn
\label{eq:basrel1a}
K_{jk}K_{kj}=1\;,\qquad
K_{jk}K_{k\ell}=K_{k\ell}K_{\ell j}=K_{\ell j}K_{jk}\;,\qquad
{[K_{jk},K_{\ell m}]}=0\;,
\nonumber\\
{[x_j,K_{k\ell}]}=
{[p_j,K_{k\ell}]}=0\;,\qquad
x_jK_{jk}=K_{jk}x_k\;,\qquad
p_jK_{jk}=K_{jk}p_k\;.
\eeqn
Thus they
are generators of a unitary representation of
the symmetric group $S_N$.
The fact that explicit realizations of these relations exist,
proves that they are internally consistent.

We will now prove that eq.~(\ref{eq:genHeisenbergalg}) follows from
the equations
(\ref{eq:basrel1}) and (\ref{eq:basrel1a}).
In fact, eq.~(\ref{eq:basrel1}) gives directly that
\beqn
{[\llaa x^n\rraa,p_j]}&=&
{[{x_j}^n,p_j]}+\sum_{k\neq j} {[{x_k}^n,p_j]}
\nonumber\\
&=&\rmi n\hbar {x_j}^{n-1}
+\rmi\lambda\hbar\sum_{k\neq j}\left(
K_{kj}{x_j}^{n-1}+x_jK_{kj}{x_j}^{n-2}+\cdots+
{x_j}^{n-1}K_{kj}\right)\nonumber\\
&\!\!\!&-\rmi\lambda\hbar\sum_{k\neq j}\left(
K_{kj}{x_k}^{n-1}+x_kK_{kj}{x_k}^{n-2}+\cdots+
{x_k}^{n-1}K_{kj}\right).
\eeqn
We observe that the $\lambda$-dependent terms
cancel, because we may convert any operator $x_k$ into $x_j$ by
pulling it through the operator $K_{kj}$.

In a similar way we have that
\beqn
{[\llaa p^n\rraa,x_j]}&=&-\rmi n\hbar{p_j}^{n-1}
-\rmi\lambda\hbar\sum_{k\neq j}\left(
K_{kj}{p_j}^{n-1}+p_jK_{kj}{p_j}^{n-2}+\cdots+
{p_j}^{n-1}K_{kj}\right)\nonumber\\
&\!\!\!&+\rmi\lambda\hbar\sum_{k\neq j}\left(
K_{jk}{p_k}^{n-1}+p_kK_{jk}{p_k}^{n-2}+\cdots+
{p_k}^{n-1}K_{jk}\right).
\eeqn
Using now also the symmetry relation $K_{jk}=K_{kj}$, we again see
that the $\lambda$-dependent terms cancel. This completes the proof.

If we use the relations
(\ref{eq:basrel1}) and (\ref{eq:basrel1a})
to compute the commutator of two symmetric
one-particle operators, the result will not in general be
a one-particle operator. It is therefore a remarkable result, proved
first in Ref.~\cite{G}, that arbitrary commutators and commutators of
commutators of the operators $\llaa x^m\rraa$ and $\llaa p^n\rraa$,
with $m,n=1,2,\ldots$, can always be written as one-particle operators.
As we have seen above, this result follows from
eq.~(\ref{eq:genHeisenbergalg}).

%%%%%%%%%%%%%%%%%%%%
\subsection{Differentiation operators}
%%%%%%%%%%%%%%%%%%%
\label{subsect:Diffop}

The index $j$ in eq.~(\ref{eq:genHeisenbergalg}) is rather superfluous,
and we may just as well write
\beqn
\label{eq:genHeisenbergalgI}
{[\llaa x^m\rraa,p]}=\rmi m\hbar x^{m-1}\;,\quad
{[\llaa p^n\rraa,x]}=-\rmi n\hbar p^{n-1}\;,\quad
{[\llaa x^m\rraa,x]}={[\llaa p^n\rraa,p]}=0\;.
\eeqn
This defines immediately a natural realization of the algebra
${\cal G}$ as a commutator algebra of differentiation operators on
the {\em non-commutative} one-particle phase space, described by
the completely non-commuting variables $x$ and $p$. Remember that $x$
and $p$ here actually represent $x_j$ and $p_j$ for one arbitrary value
of the index $j$, and in general we have no commutation rules for
reordering monomials in $x$ and $p$.

A linear operator ${\cal A}:f\mapsto {\cal A}(f)$ is called
a differentiation operator if it acts according to the Leibniz rule,
\beqn
{\cal A}(fg)={\cal A}(f)g+f{\cal A}(g)\;.
\eeqn
The commutator ${\cal C}={[{\cal A},{\cal B}]}$ of two differentiation
operators ${\cal A}$ and ${\cal B}$ is defined in the obvious way,
\beqn
{\cal C}(f)={\cal A}({\cal B}(f))-{\cal B}({\cal A}(f))\;.
\eeqn
It is easily verified that ${\cal C}$ is again a differentiation
operator.

Any operator ${\cal A}$ of the form
${\cal A}:f\mapsto{[A,f]}$
is automatically a differentiation operator. Moreover, the mapping
$A\mapsto{\cal A}$ preserves the commutator product. In fact, if
${\cal A}(f)={[A,f]}$ and ${\cal B}(f)={[B,f]}$, then the commutator
${\cal C}={[{\cal A},{\cal B}]}$ is given by
the Jacobi identity as
\beqn
{\cal C}(f)={[A,\,[B,f]\,]}-{[B,\,[A,f]\,]}={[\,[A,B]\,,f]}\;.
\eeqn

Due to the Leibniz rule, the action of a differentiation operator on
any
polynomial
in the (commuting or non-commuting) variables $x$ and $p$ is uniquely
given by its action on $x$ and $p$.
Let ${\cal X}_m$ and ${\cal P}_n$ be the differentiation operators
representing $\llaa x^m\rraa$ and $\llaa p^n\rraa$, defined such that
\beqn
\label{eq:basicdiffops}
{\cal X}_m(f)={[\llaa x^m\rraa,f]}\;,\qquad
{\cal P}_n(f)={[\llaa p^n\rraa,f]}\;.
\eeqn
Then, as an example, the commutator
$A={[\llaa x^m\rraa,\llaa p^n\rraa]}$
is represented by the differentiation operator
${\cal A}={[{\cal X}_m,{\cal P}_n]}$,
acting in the following way,
\beqn
\label{Apx}
{\cal A}(x)&=&
{\cal X}_m({\cal P}_n(x))=
-\rmi n\hbar{\cal X}_m(p^{n-1})=
mn\hbar^2\sum_{k=0}^{n-2}p^kx^{m-1}p^{n-2-k}\;,\nonumber\\
{\cal A}(p)&=&
-{\cal P}_n({\cal X}_m(p))=
-\rmi m\hbar{\cal P}_n(x^{m-1})=
-mn\hbar^2\sum_{k=0}^{m-2}x^kp^{n-1}x^{m-2-k}\;.\;\;\;\;
\eeqn

A general element of the algebra ${\cal G}$
can thus be represented
explicitly
in the form
\be
{\cal A}=A_x\,\pdd{}{x}+ A_p\,\pdd{}{p}\;,
\label{A-unique}
\ee
showing how the operator ${\cal A}$ acts on $x$ and $p$.
Here
$A_x={\cal A}(x)$ and  $A_p={\cal A}(p)$
are polynomials in $x$ and $p$.
We stress that the representation (\ref{A-unique}) is {\em unique}.

%%%%%%%%%%%%%%%%%%%%%%%
\subsection{The
Hermitian matrix model}
%%%%%%%%%%%%%%%%%%%%%%%%%
\label{subsect:clmatmod}

In this model there are $N^2$ real coordinates arranged into
an $N\times N$ complex Hermitian matrix $X_{jk}$. The $N^2$
conjugate momenta are similarly arranged into a Hermitian matrix
$P_{jk}$. The Hermitian symmetry conditions for the classical 
variables,
$X_{jk}^{\ast}=X_{kj}$ and $P_{jk}^{\ast}=P_{kj}$, are replaced by
the conditions $X_{jk}^{\dag}=X_{kj}$ and $P_{jk}^{\dag}=P_{kj}$
for the operators in the quantum theory. The Poisson brackets and
the corresponding commutation relations are the following,
\beqn
\label{eq:commrelmatrix}
\{X_{jk},P_{\ell m}\}=\delta_{jm}\delta_{k\ell}\;,\qquad
{[X_{jk},P_{\ell m}]}=\rmi\hbar\delta_{jm}\delta_{k\ell}\;,
\eeqn
and they are compatible with the Hermitian symmetry of the matrices
$X$ and $P$.

In order to relate this matrix model to the theory of a system of
identical particles, we interpret the special unitary group
$\mbox{SU}(N)$ as an extension of the
symmetric group $S_N$ which interchanges the particle positions.
An $\mbox{SU}(N)$ matrix $U$ acts on the matrices $X$ and $P$ by
conjugation, as
\beqn
X\mapsto UXU^{-1}\;,\qquad
P\mapsto UPU^{-1}\;.
\eeqn
The Poisson brackets or commutation relations,
eq.~(\ref{eq:commrelmatrix}), are preserved.
The Hermitian matrix
$X$ may always be diagonalized by
an $\mbox{SU}(N)$ matrix $U$,
and the
eigenvalues $x_1,x_2,\ldots,x_N$
may be interpreted as the particle positions
on the line.
Different $\mbox{SU}(N)$ matrices which
diagonalize $X$ may give different orderings of the
eigenvalues.
The full $\mbox{SU}(N)$ group now
gives a {\em continuous} way to interchange the particle
positions \cite{PolyPLB}.

The observables of the $N$-particle system correspond to the
$\mbox{SU}(N)$ invariant quantities of the matrix model. Thus, the
observable part of the
$N\times N$ Hermitian matrix $X$ is its set of $N$ real
eigenvalues,
in arbitrary order. An equivalent complete set of
$\mbox{SU}(N)$ invariants for the matrix $X$ are the symmetric
one-particle observables
\beqn
\llaa x^m\rraa=\sum_{j=1}^N{x_j}^m=\tr X^m\;,
\eeqn
with $m=1,2,\ldots,N$. Thus we see that the observables in
the Hermitian matrix model which correspond to one-particle
observables for a system of identical particles, such as
$\llaa x^m\rraa$, $\llaa p^n\rraa$, $\llaa x^mp^n\rraa$,
are traces of matrix products: $\tr X^m$, $\tr P^n$,
and in general,
\beqn
\label{eg:f}
f=\tr(F_1F_2\cdots F_M)\;,
\eeqn
where each matrix $F_j$ is either $X$ or $P$.
We say that $f$ has order $(m,n)$ if the matrix product contains
$m$ factors $X$ and $n$ factors $P$.
This is a realization of the defining
relations in eq.~(\ref{eq:genHeisenbergalg}),
in the sense that, e.g.,
\beqn
\label{eq:commrelmatrixII}
{[\tr X^m,P_{jk}]}=\rmi m\hbar\,(X^{m-1})_{jk}\;,\qquad
{[\tr P^n,X_{jk}]}=-\rmi n\hbar\,(P^{n-1})_{jk}\;.
\eeqn

In the classical matrix model we are free to permute cyclically
the factors of a matrix product inside the trace, so that
$f=\tr(F_2\cdots F_MF_1)=\ldots$.
In the quantized theory the same relations hold modulo
correction terms proportional to powers of $\hbar$, due
to the Heisenberg commutation relations.
Note, however, that since the basic commutation relations 
(\ref{eq:commrelmatrixII}) hold equally well both in the classical
and the quantum case, the algebra ${\cal G}$ generated is the same in
both cases. Therefore, we can consider for simplicity the classical
case, ignoring correction terms.  The classical matrix
model is able to embody the {\it full} structure of the 
algebra ${\cal G}$
(unlike the classical $N$-particle phase space), since the (quantum)
non-commutativity of $x$ and $p$, postulated in 
(\ref{eq:genHeisenbergalg}), is replaced by
the classical non-commutativity of matrices $X$ and $P$.

The set of all linear combinations of terms
of the form $f$, eq.~(\ref{eg:f}), is closed under
Poisson brackets and is therefore a Lie algebra of
symmetric one-particle observables. Let us call it ${\cal F}$.
Note that the element $f\in{\cal F}$ is in general complex,
and its complex conjugate is
\beqn
f^{\ast}=\tr((F_1F_2\cdots F_M)^{\dag})=\tr(F_M\cdots F_2F_1)\;.
\eeqn
A subalgebra of ${\cal F}$ is the set ${\cal F}_R$ of all
linear combinations of terms of the form $f+f^{\ast}$.

We can use the observables $\tr X^m$, $\tr P^n$ to generate
the algebra
${\cal G}$ as a subalgebra of ${\cal F}_R$.
For example, considering the commutator
which had an ambiguous representation in the equations
(\ref{xp1}) and (\ref{xp2}),
we have
\beqn
\{\tr X^m,\tr P^n\}=mn\tr\,({X}^{m-1}{P}^{n-1})\;.
\label{cyclic}
\eeqn
This Poisson bracket can be computed by means of
the canonical relations (\ref{eq:commrelmatrix}),
or in two other ways by (the Poisson bracket version
of) eq.~(\ref{eq:commrelmatrixII}).
The result must be independent of the method, since there can
be no reordering problems in the classical case.
Thus there is no ambiguity
in the representation of the Poisson bracket (\ref{cyclic})
as an observable of the classical Hermitian matrix model.
It follows in a similar way that every element of ${\cal G}$
has a unique representation in this model.
For this reason, the matrix model realization of ${\cal G}$
is perhaps the most efficient and convenient of all
realizations, for computational purposes.

As already pointed out, the $N$-particle system can be considered as 
corresponding to the $\mbox{SU}(N)$ invariant  part of the matrix 
model. The matrix model is larger since it also includes  
$\mbox{SU}(N)$-dependent, 
``angular'' variables. Quantum mechanically, different 
SU($N$) ``angular momentum'' representations can be given by 
specifying values for certain constants of motion. In this way 
different $N$-particle 
systems can be constructed, such as free fermions, the  
Calogero system \cite {OlPerPhRep}, and  Calogero-like systems 
with additional internal degrees of freedom \cite{MinPolPLB}. 
It is of interest to note that for all these systems, 
${\cal G}$ is the relevant algebra of $\mbox{SU}(N)$ invariant  
observables.
%%%%%%%%%%%%%%%%%%%%%%%%%%%%%%%%
\section{Remarks on the structure of
${\cal G}$}
\label{structure}
%%%%%%%%%%%%%%%%%%%%%%%%%%%%%%%%

In this section we present results of computer (symbolic)
calculations using the realizations of the  algebra ${\cal G}$
given in the two previous subsections.
We have computed the degeneracy $g(m,n)=g(n,m)$,
i.e.\ the number of linearly independent elements of ${\cal G}$
of order $(m,n)$, for  $m+n\leq 16$, as presented in Figure~1.
In Ref.~\cite{G} the degeneracies were given for $m+n\leq 6$.

Let us outline how to compute using the classical
Hermitian matrix model. One need not use all the observables
$\tr X^m$ and $\tr P^n$ to generate ${\cal G}$ as described in
Sect.~\ref{definition}, 
it is enough to use only a finite set
of generators, e.g.\ $\tr P$, $\tr X^2$, and $\tr P^3$
\cite{G}.
One may generate all of ${\cal G}$ by starting with $\tr P$
and proceeding as follows.
All the observables of order $(m,n)$ are obtained
from those of order $(m-1,n+1)$ and $(m+1,n-2)$,
by means of Poisson brackets with $\tr X^2$
and $\tr P^3$, respectively.
Linear dependence of the generated observables
of order $(m,n)$ has to be checked.

Defining the partial generating functions
\be
G_{n}(q)=\sum_{m=0}^{\infty} g(n,m)\, q^m \;,
\label{Gn}
\ee
we recover the degeneracies $g(n,m)=g(m,n)$ listed in Fig.~1
for $n\leq 3$ with
\be
\label{G123}
G_0&=&G_1=\frac{1}{1-q} \;, \nonumber\\
G_2&=&\frac{1}{(1-q)(1-q^2)} \;, \\
G_3&=&\frac{1}{(1-q)(1-q^2)(1-q^3)} \;. \nonumber
\ee
This formula for the generating functions $G_0$ and $G_1$
is easily proved to all orders. For $G_2$ and $G_3$ we have no proof,
only numerical verification up to the level $m+n=16$.
Eq.~(\ref{G123}) would imply that for $n\leq 3$,
\be
g(n,m)=p_n(m) \;,
\label{g}
\ee
the number of partitions of $m$ into at most $n$ parts.
$p_n(m)$ is in fact the degeneracy for order $(m,n)$ with $n\leq 3$
of the algebra ${\cal F}_R$, which was defined in
Subsect.~\ref{subsect:clmatmod} and which contains ${\cal G}$.
To see this in the case $n=3$, as an example, note that the general 
basis element of ${\cal F}_R$ then has the form
\be
\tr(X^{m_1}PX^{m_2}PX^{m_3}P+PX^{m_3}PX^{m_2}PX^{m_1})\;,
\ee
where $m_1,m_2,m_3$ are non-negative integers and
$m_1+m_2+m_3=m$. This trace is invariant under any permutation
of $m_1,m_2,m_3$, showing that there is a one to one
correspondence between basis elements of order $(m,3)$ and
partitions of $m$ into at most three parts.

Since ${\cal G}$ is a subalgebra of ${\cal F}_R$, it is a remarkable
fact that these two algebras are actually identical for every
order $(m,n)$ with $m+n\leq 16$ and either $m\leq 3$ or $n\leq 3$.
Our conjecture is that this is true for arbitrary $m$ and $n\leq 3$,
or equivalently, for arbitrary $n$ and $m\leq 3$.

In contrast, for $m\geq 4$ and $n\geq 4$ we find that
${\cal F}_R$ is larger than ${\cal G}$.
We may cite here two more generating functions,
\be
G_4&=&\frac{1+q^2}{(1-q)(1-q^2)(1-q^3)(1-q^4)}\;,\nonumber\\
G_5&=&\frac{(1+q^2)(1+q^3+q^4)}{(1-q)(1-q^2)(1-q^3)(1-q^4)(1-q^5)}\;,
\label{G45}
\ee
which recover in a non-trivial way the degeneracies
in Fig.~1 for $n=4$ and $n=5$, respectively.
{From} these expressions as well as from Fig.~1 one can see that
the degeneracies $g(m,n)$ of ${\cal G}$ for $m\geq 4$ and $n\geq 4$
are larger than  the numbers of partitions $p_n(m)$.

\section{Conclusions}

We have presented a formulation of the
algebra of observables for identical particles on a line
in terms of a previously studied infinite dimensional
Lie algebra ${\cal G}$, starting from a set of
basic commutation relations which are parameter-independent.
We have given two new realizations of
${\cal G}$ in which the elements are represented
without ambiguities.
This provides a simpler way, both conceptually and practically,
to study the structure of the algebra.
The advantages of the new representations were demonstrated
by evaluating the
degeneracies of
${\cal G}$ for low orders.

The algebra  ${\cal G}$ is overcomplete with respect to the classical 
coordinates and momenta of the $N$-particle system, and this is 
exposed by the degeneracies for given order $(m,n)$ of the algebra,
which we have examined in this paper. 
An obvious interpretation of these degeneracies   
 is that the algebra, in general, involves more degrees of freedom  
than those present in  the classical $N$-particle system. 
(Note, however, that for irreducible representations of ${\cal G}$,
the number of independent variables will be smaller than indicated 
by the degeneracies since the presence of Casimir operators 
will introduce identities between elements of the algebra.)  
The conceptual questions then remain: 
What is the most general class
of particle systems for which ${\cal G}$ embodies the full
set of observables, and can the additional variables be related to  
internal degrees of freedom for such systems? 
Answers to these questions depend on the nature of unitary irreducible 
representations of ${\cal G}$,  which are yet to be studied. 

\bigskip
S.B.I. and R.V. acknowledge support from the Norwegian Research Council.

%%%%%%%%%%%%%%%%%%%%%%%%%%%%%%%% BIBLIOGRAPHY
%%%%%%%%%%%%%%%%%%%%%%%%%%%%%%%%%%%%%%%%%

\begin{figure}
\epsfxsize=6.0in
\epsffile{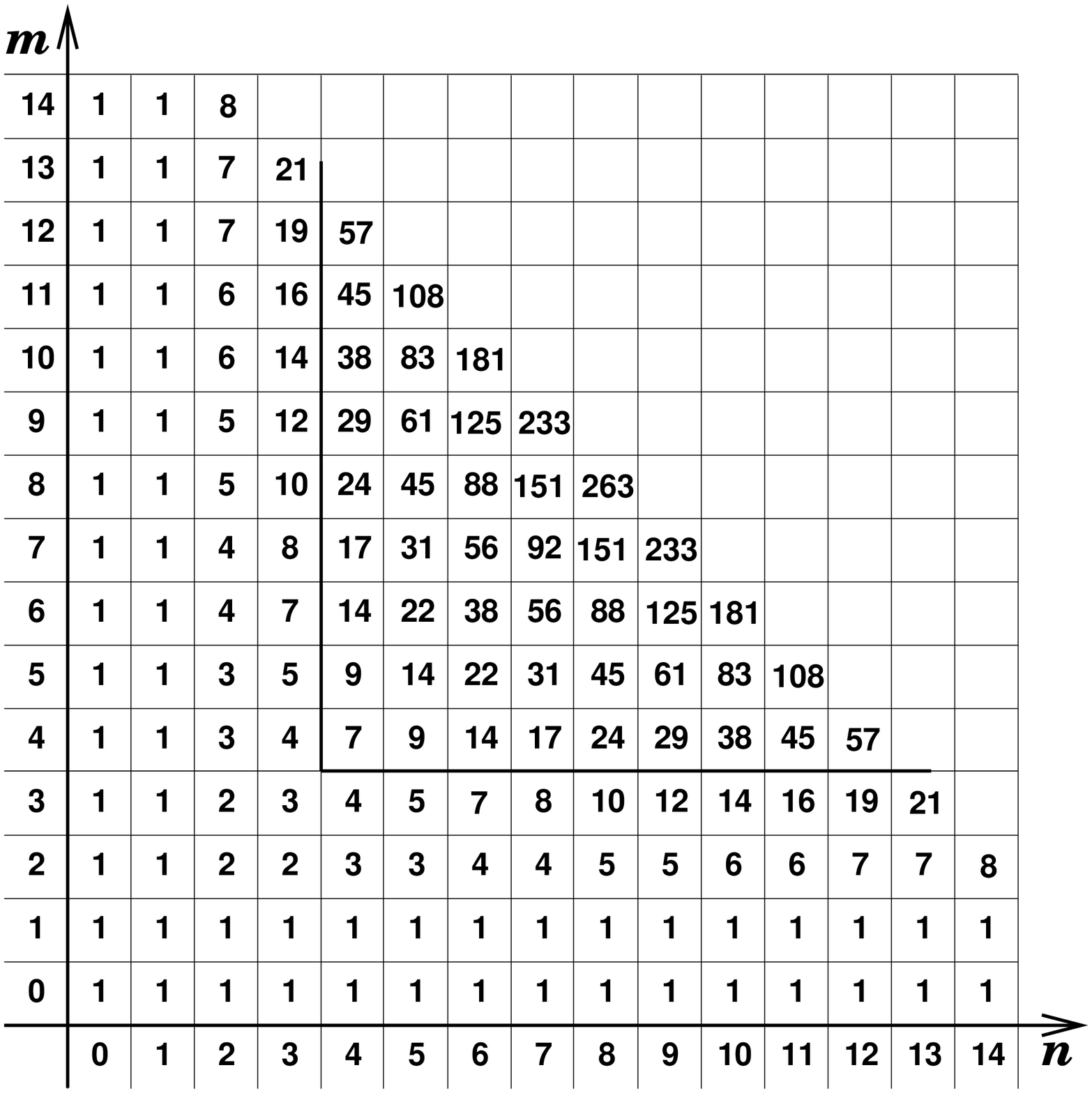}
\caption{The
degeneracy $g(m,n)$ of the algebra ${\cal G}$
as a function of the order $(m,n)$.
For ${\rm min}(m,n)\leq 3$, the degeneracies
equal
the numbers of partitions of ${\rm max}(m,n)$ into at most
${\rm min}(m,n)$ parts. For ${\rm min}(m,n)> 3$, the degeneracies are
larger than the numbers of
partitions.  }
\end{figure}


\newpage
\begin{thebibliography}{99}


\bibitem{WHPAMD} W. Heisenberg,
Zeits. f. Phys. {\bf 38} (1926) 411;\\
P. A. M. Dirac, Proc. Roy. Soc. London (A) {\bf 112} (1926) 661;\\
{\em The Principles of Quantum Mechanics},
Oxford University Press (1935).

\bibitem{LM-Heis} J. M. Leinaas and J. Myrheim,
\journal Phys. Rev. B, 37, 9286, 1988;\\
\journal Int. J. Mod. Phys. B, 5, 2573, 1991;
\journal Int. J. Mod. Phys. A, 8, 3649, 1993.

\bibitem{Cal} F. Calogero, J. Math. Phys. {\bf 10} (1969) 2191, 2197;
{\bf 12} (1971) 419.

\bibitem{Perelomov} A. M. Perelomov, Teor. Mat. Fiz. {\bf 6} (1971) 
364.

\bibitem{PolyNPB} A. P. Polychronakos, \journal Nucl. Phys. B, 324,
597, 1989.

\bibitem{HLM} T. H. Hansson, J. M. Leinaas, and J. Myrheim,
\journal Nucl. Phys. B, 384, 559, 1992.

\bibitem{I-IJMPA}  S. B. Isakov, \journal Int. J. Mod. Phys. A, 9, 
2563, 1994.

\bibitem{Haldane} F. D. M. Haldane,
\journal Phys. Rev. Lett., 67, 937, 1991.

\bibitem{I-MPLB}  S. B. Isakov,
\journal Mod. Phys. Lett. B, 8, 319, 1994.

\bibitem{Wu94} Y. S. Wu, \journal Phys. Rev. Lett., 73, 922, 1994.

\bibitem{dVO94}  A. Dasni{\`e}res de Veigy and S. Ouvry,
\journal Phys. Rev. Lett., 72, 600, 1994;


\bibitem{Poly-PRL92} A. P. Polychronakos,
Phys. Rev. Lett. {\bf 69} (1992) 703.

\bibitem{BHV} L. Brink, T. H. Hansson, and M. Vasiliev,
Phys. Lett. B {\bf 286} (1992) 109;\\
L. Brink, T. H. Hansson, S. E. Konstein, and M. Vasiliev,
Nucl. Phys. B {\bf 384} (1993) 591.

\bibitem{G} S. B. Isakov and J. M. Leinaas,
Nucl. Phys. B {\bf 463} (1996) 194.

\bibitem{PolyPLB} A. P. Polychronakos, Phys. Lett. B {\bf 266} 
(1991) 29.

\bibitem{OlPerPhRep} M. A. Olshanetskii and A. M. Perelomov, Phys.
Rep. {\bf 94} (1983) 313.

\bibitem{MinPolPLB} J. Minahan and A. P. Polychronakos, Phys. Lett. B 
{\bf 326} (1994) 288.

\end{thebibliography}
\end{document}